\documentclass[11pt,a4paper]{article}

\usepackage[utf8]{inputenc}
\usepackage[T1]{fontenc}
\usepackage[margin=2.5cm]{geometry} 
\usepackage{authblk}               
\usepackage{hyperref}               
\usepackage{amsmath, amssymb}       
\usepackage{graphicx}               
\usepackage{subcaption}

\hypersetup{
    colorlinks=true,
    linkcolor=blue,
    filecolor=magenta,
    urlcolor=cyan,
    pdftitle={Titolo del tuo paper},
    pdfpagemode=FullScreen,
}

\usepackage{fancyhdr}
\fancypagestyle{firstpage}{
    \fancyhf{} 
    \fancyfoot[C]{\footnotesize Preprint version. Accepted for publication in \textit{Statistics} (Taylor \& Francis). \\ Final version available at \url{https://doi.org/10.1080/02331888.2026.2636122}}
}

\title{Statistical Measures for Explainable Aspect-Based Sentiment Analysis: A Case Study on Environmental Discourse in Reddit}

\author[1]{Luisa Stracqualursi}
\author[2]{Patrizia Agati}
\affil[1,2]{University of Bologna, Departments of Statistics,
via Belle Arti 41, 40126 Bologna, Italy}

\date{} 

\begin{document}

\maketitle              
\thispagestyle{firstpage}

\begin{abstract}
Aspect-Based Sentiment Analysis (ABSA) provides a fine-grained understanding of opinions by linking sentiment to specific aspects in text. While transformer-based models excel at this task, their black-box nature limits their interpretability, posing risks in real-world applications without labeled data. This paper introduces a statistical, model-agnostic framework to assess the behavioral transparency and trustworthiness of ABSA models. Our framework relies on several metrics, such as the entropy of polarity distributions, soft-count-based dominance scores, and sentiment divergence between sources, whose robustness is validated through bootstrap resampling and sensitivity analysis. A case study on environmentally focused Reddit communities illustrates how the proposed indicators provide interpretable diagnostics of model certainty, decisiveness, and cross-source variability. The results show that statistical indicators computed on soft outputs can complement traditional approaches, offering a computationally efficient methodology for validating, monitoring, and interpreting ABSA models in contexts where labeled data are unavailable.
\end{abstract}


\section{Introduction}
In recent years, the interpretability of machine learning models has become a key concern across a wide range of applications, especially in domains where transparency, accountability, and trust are critical. This concern has led to the emergence of eXplainable Artificial Intelligence (XAI), a research field that aims to develop methods capable of making model decisions understandable to human users \cite{lipton2018,yang2023}.
The challenge is particularly pronounced for deep learning models—such as transformers—due to their high capacity, opacity, and sensitivity to training data \cite{Rudin2019}.

Aspect-Based Sentiment Analysis (ABSA) presents an ideal testbed for explainability research. As a fine-grained extension of traditional sentiment analysis, ABSA involves detecting opinion-bearing aspects within a sentence and classifying the associated sentiment polarity \cite{Pontiki2014}. The ability to attribute sentiment not just to entire documents but to individual aspects allows for a more transparent interpretation of user opinions, making ABSA particularly relevant in domains such as product reviews, political debate, or public discourse.

With the rise of transformer-based language models, ABSA has seen major improvements in performance \cite{Xu2020}, though these gains often come at the cost of reduced transparency.
Several XAI methods have been proposed to address this, including attention-based visualization \cite{Vig2019}, Local Interpretable Model-agnostic Explanations (LIME) \cite{ribeiro2016}, SHapley Additive exPlanations (SHAP) \cite{Lundberg}, Integrated Gradients \cite{Sundararajan2017}, and model-specific methods for attention attribution. While effective, these techniques often require access to model internals or rely on costly post-hoc computations, limiting their applicability in real-world, noisy settings.

In response to these limitations, this paper shifts the focus from traditional accuracy-based evaluation, which relies on labeled data and fine-tuning, to a statistical analysis of model behavior on data without ground-truth labels. We propose a model-agnostic, computationally efficient framework that utilizes a suite of metrics to assess the behavioral properties of a pretrained ABSA model. Drawing inspiration from the SAFE paradigm \cite{Giudici2022,Giudici2023,Giudici2025,BabaeiGiudici2025}—Sustainable, Accurate, Fair, and Explainable machine learning—we introduce metrics to gauge polarity uncertainty (entropy), polarity assertiveness (dominance), and sentiment divergence between different sources.

To ground our approach in a practical application, we apply it to a case study involving Reddit discussions from environmentally focused communities: r/environment, r/climatechange, and r/renewableenergy. These platforms represent a rich, organically generated corpus of public opinion on energy and sustainability topics. Our findings demonstrate that this statistical framework can yield interpretable and actionable insights on ABSA behavior, even in the absence of ground-truth labels or task-specific fine-tuning.

The main contributions of this work are centered on providing a low-cost framework that leverages unlabeled data to address three critical challenges in the Machine Machine Learning Operations (MLOps) lifecycle for ABSA models:
\begin{enumerate}
  \item Pre-deployment Behavioral Validation. While traditional validation relies on labeled test sets to measure accuracy, our framework enables a complementary form of pre-deployment assessment. By analyzing metrics such as output stability via bootstrap resampling and prediction certainty via entropy, practitioners can perform crucial "sanity checks" on a model's behavior. This allows for the early detection of inherent instabilities or problematic biases, enabling teams to assess a model's fundamental suitability for a domain before committing resources to a full-scale annotation and fine-tuning project.

  \item Intelligent Prioritization of Annotation Efforts. The process of data annotation is often a significant bottleneck due to its high cost. Our framework can be applied to large, unlabeled corpora to identify data subsets where the model exhibits the most uncertainty (e.g., high entropy or high source divergence). This functions as a form of active learning, guiding annotation teams to focus their resources on the most informative and challenging samples. This targeted approach maximizes the efficiency of the annotation budget and enhances the effectiveness of subsequent fine-tuning stages.

  \item Post-deployment Monitoring for Model Drift. Once deployed, Natural Language Processing (NLP) models are susceptible to performance degradation over time as the distribution of real-world data changes—a phenomenon known as model drift. Continuous re-annotation for monitoring is often unfeasible. Our framework offers a practical solution by enabling the time-series tracking of the model's output statistics. A significant shift in these metrics can serve as a reliable proxy indicator of a potential drift, providing an early-warning system that triggers investigation and model retraining cycles, thus ensuring long-term reliability.
\end{enumerate}

\section{Related Work}

Research on explainability in Aspect-Based Sentiment Analysis (ABSA) has evolved along several distinct lines. The most common approach involves \emph{post-hoc explanation methods}, which apply local interpretability techniques to transformer models. Perikos and Diamantopoulos (2024), for example, utilized LIME, SHAP, and gradient-based methods to interpret fine-tuned ABSA models at the token level \cite{Perikos2024}. Their findings confirm that these techniques can identify influential input features but also highlight significant limitations, including high computational costs and sensitivity to model architecture.

A second stream of research focuses on \emph{hybrid and inherently interpretable models}, which aim to embed interpretability directly into the model's design. This aligns with calls for greater transparency in high-stakes decision-making \cite{Rudin2019}. Approaches in this area, such as SA-EXAL \cite{Pereg2020}, often integrate syntactic and linguistic structures into attention mechanisms to enhance the plausibility of the explanations.

More aligned with our work are \emph{global and statistical approaches}, which explain model behavior by aggregating predictions at a corpus or aspect level rather than interpreting them individually. While methods like \emph{influence functions }\cite{KohLiang2017} and \emph{concept activation vectors} \cite{Kim2018} have been used to assess model stability, their application to ABSA in unstructured, non-annotated contexts remains limited. A notable contribution in this stream is the SAFE (Sustainable, Accurate, Fair, and Explainable) framework, which introduces a suite of statistical indicators to quantify principles such as decisional transparency and model stability. While the SAFE framework provides a solid theoretical foundation, the direct application of its principles to the outputs of ABSA models in the absence of labeled data is an area that has, until now, remained largely unexplored.

Our work, therefore, contributes to this line of statistical research by addressing these specific gaps. We propose a tailored framework of lightweight, model-agnostic metrics (entropy, sentiment dominance, and divergence) designed to operate effectively on ABSA predictions in large-scale, naturally occurring text corpora, as demonstrated through our case study.

\section{Methodology}

Our methodology introduces a statistical, model-agnostic framework to assess the behavioral properties of a pretrained ABSA model. In this context, our approach focuses on operationalizing the principles of Explainability and Sustainability through computationally efficient metrics that do not require labeled data. The methodology proceeds in three main stages: (1) aspect-based sentiment prediction to obtain probabilistic outputs; (2) computation of statistical indicators for explainability; and (3) a robustness assessment of these indicators.

\subsection{Stage 1: Aspect Extraction and Probabilistic Sentiment Prediction}
 To conduct ABSA, we employed the Python implementation proposed by Yang and Chen (2022), namely pyABSA, an open-source library built on PyTorch \cite{paszke2019pytorch,yang2023}. We used the FASTATPC model, which is available through PyABSA (v2.4.1.post1), for automatic aspect extraction and sentence-level sentiment classification \cite{qin2021pyabsa}. Crucially, its default prediction routine returns not only the predicted polarity and confidence score for each instance but also the full softmax probability distribution across sentiment classes (Positive, Negative, Neutral). This capability allows us to compute probabilistic explainability metrics without modifying the model architecture, providing the soft-label outputs essential for quantifying uncertainty.

\subsection{Stage 2: Computation of Statistical Indicators} \label{par2.2}

From the probabilistic outputs, we first define a sentiment profile for each extracted aspect $A$. This profile, $\tilde{w}(A,s)$, represents the
expected probability that aspect $A$ is associated with sentiment $s \in \{\textit{Positive}, \textit{Negative}, \textit{Neutral}\}$ across all its $N_A$ occurrences in the corpus:

\begin{equation}\label{1}
\tilde{w}(A, s) = \frac{1}{N_A} \sum_{i \in \mathcal{I}_A} p_i(s),
\end{equation}
where $\mathcal{I}_A$ is the set of all instances where aspect $A$ appears,
$N_A = |\mathcal{I}_A|$, and $p_i(s)$ is the model-assigned probability for
sentiment $s$ in instance $i$.

It is important to note that an "aspect" term may occur multiple times across different sentences, each associated with potentially different sentiment
polarities. Rather than collapsing these into a single majority label, our approach aggregates the full probability distributions to build a soft
sentiment profile $\tilde{w}(A,s)$. This representation retains both the dominant polarity and the variability of user perceptions, thus preserving the model's uncertainty.

\vspace{0.8em}
\paragraph{Illustrative Example.}\label{esempio1} Consider the aspect \textit{battery}, which occurs in three sentences with the following distributions:

\begin{itemize}
  \item Sentence 1: \{Positive: 0.91, Negative: 0.06, Neutral: 0.03\}
  \item Sentence 2: \{Positive: 0.08, Negative: 0.87, Neutral: 0.05\}
  \item Sentence 3: \{Positive: 0.21, Negative: 0.19, Neutral: 0.60\}
\end{itemize}

The aggregated profile is:
\[
\tilde{w}(\textit{battery}, s) =
\{ \textit{Positive}: 0.40,\ \textit{Negative}: 0.37,\ \textit{Neutral}: 0.23 \}.
\]

This example shows how probabilistic aggregation captures not only the dominant polarity but also the strength of the predictions. In
the next step, we introduce two complementary indicators --- \emph{polarity entropy} and \emph{dominance} --- to quantify these distributional properties of
$\tilde{w}(A,s)$.

\paragraph{Polarity Entropy $H(A)$.} It quantifies the uncertainty in an aspect's sentiment distribution. Higher values indicate more ambiguous sentiment profiles, while lower values suggest decisional clarity. It is computed using the Shannon Entropy formula, which for a discrete probability distribution is:

\begin{equation}\label{2}
H(A) = - \sum_{s} \tilde{w}(A, s) \cdot \ln(\tilde{w}(A, s))
\end{equation}

This formula uses the natural logarithm (ln), meaning the unit of measure is \emph{nats}. With three sentiment classes (Positive, Negative, Neutral), the entropy value H(A) ranges from a minimum of 0 (for a certain prediction) to a maximum of $\ln 3 \approx 1.0986$ (for maximum uncertainty, i.e., a uniform distribution).

Eq. \ref{2} represents the "pure", or unnormalized, entropy. To obtain a normalized entropy, which ranges from 0 to 1, this value can be divided by the maximum possible entropy for three classes (Positive, Negative, Neutral), which is $\ln 3$.

Revisiting the example above (example \ref{esempio1}), where the model exhibits a very high uncertainty regarding the aspect's sentiment, we obtain:

\begin{itemize}
    \item $H(\textit{battery}) = - (0.40 \ln 0.40 + 0.37 \ln 0.37 + 0.23 \ln 0.23) \approx 1.0724$
\end{itemize}

\paragraph{Sentiment Dominance $D(A)$.} It quantifies the strength of the dominant sentiment, taking values from $1/3$ (maximum uncertainty) to $1$ (certainty), and is defined as:

\begin{equation}\label{3}
D(A) = \max_{s} \tilde{w}(A, s)
\end{equation}

It is worth noting that Dominance is distinct from the indicator known as
\emph{Confidence} in the context of ABSA models. When multiple instances are considered, this distinction is formally
captured by the inequality:
\[
E[\max_s p_i(s)] \geq \max_s E[p_i(s)].
\]
This relationship is a direct consequence of Jensen's inequality applied to
the convex $\max(\cdot)$ function. The two terms correspond, respectively, to the
definitions of Confidence and Dominance.

\paragraph{Formal Definitions.}
\begin{itemize}
    \item \textbf{Confidence:} ${Conf}(A) = E[\max_s p_i(s)]$.
    Confidence is the expected value of the maximum prediction probability
    across sentiment classes $s$ for a given instance $i$. It is obtained by
    identifying, for each instance, the highest probability among the sentiment
    classes and then averaging these maxima. This metric reflects the model's
    average conviction in its per-instance predictions, regardless of which
    sentiment is predicted.

    \item \textbf{Dominance:} $ D(A) = \max_s E[p_i(s)]$.
    Dominance is defined as the maximum of the expected probability values for
    each sentiment class. It is obtained by first averaging the probability of
    each class across all instances, and then selecting the largest among these
    averages. This metric measures the overall prevalence of a sentiment across
    the dataset for a given aspect.
\end{itemize}

\paragraph{Illustrative Example.}
Consider the aspect \emph{climate} evaluated over two instances. Let
$p_i(s)$ denote the probability assigned by the model to sentiment $s$ for
instance $i$.

\begin{table}[htbp]
\centering
\scriptsize
\begin{tabular}{lcccc}
\hline
Instance & $p(\textit{positive})$ & $p(\textit{neutral})$ & $p(\textit{negative})$ & $\max_s p_i(s)$ \\
\hline
 1 & 0.90 & 0.05 & 0.05 & 0.90 \\
 2 & 0.10 & 0.10 & 0.80 & 0.80 \\
\hline
Expectation ($E$) & 0.50 & 0.075 & 0.425 & -- \\
\hline
\end{tabular}
\caption{Example: probabilities for aspect \emph{climate} across two istances.}
\end{table}

\textbf{Dominance Calculation.}
The expected probabilities are:
$E[p(\textit{positive})]=0.50$,
$E[p(\textit{neutral})]=0.075$,
$E[p(\textit{negative})]=0.425$.
The maximum of these is $\max(0.50, 0.075, 0.425) = 0.50$,
so the Dominance score is $D(A)=0.50$ (dominant sentiment: positive).

\textbf{Confidence Calculation.}
The maximum probability for each instance is 0.90 (Instance~1) and 0.80
(Instance~2). Their mean is
\[
E[\max_s p_i(s)] = \frac{0.90+0.80}{2} = 0.85.
\]
Thus, the Confidence is 0.85.

As predicted by the inequality, Confidence (0.85) exceeds Dominance (0.50).
This illustrates that Confidence reflects the model’s per-instance certainty,
whereas Dominance captures the aggregate sentiment trend across instances.

\paragraph{Sentiment Divergence $JSD(A)$.}
To quantify how the sentiment expressed towards an aspect $A$ differs between two distinct data subsets, we employ the well known Jensen-Shannon Divergence (JSD) \cite{Lin1991}. The JSD is a symmetric and bounded measure of dissimilarity between two probability distributions, making it ideal for comparing sentiment profiles. Given the sentiment profiles for an aspect $A$ calculated on two separate probability distributions, $\tilde{w}_1$ and $\tilde{w}_2$, the divergence is computed as:

\begin{equation}\label{4}
JSD(A) = \frac{1}{2} D_{KL}(\tilde{w}_1 \parallel M) + \frac{1}{2} D_{KL}(\tilde{w}_2 \parallel M)
\end{equation}

where $M = \frac{1}{2}(\tilde{w}_1 + \tilde{w}_2)$  is a mixture distribution of $\tilde{w}_1$ and $\tilde{w}_2$, and $D_{KL}$ denotes the Kullback--Leibler divergence. The JSD score ranges from $0$, indicating that the sentiment distributions are identical, to $1$, indicating they are entirely different.

The subsets for this comparison can be defined based on various criteria relevant to the research question. For instance, they can represent different conversational roles within a community (e.g., original posts vs.\ comments) or different communities altogether (e.g., one subreddit vs.\ another). The specific application of this metric is detailed in our case study section.

\subsection{Stage 3: Robustness and Stability Assessment} \label{par:2.3}
To evaluate the stability of our metrics, we implement two complementary procedures.

\paragraph{Confidence-Based Filtering and Robustness Assessment.}
To assess the robustness and stability of aspect-level sentiment profiles under uncertainty, we implement a confidence-based filtering procedure inspired by the Rank Graduation Robustness (RGR) principle in the SAFE framework.

For each prediction $i$, the classifier returns a confidence score $C(A)_i \in [0, 1]$. By discarding instances below increasing confidence thresholds $\theta \in \{0.0, 0.2, 0.4, 0.6, 0.8\}$, we simulate an explainability--robustness trade-off: as more uncertain predictions are excluded, we expect:

\begin{itemize}
    \item a decrease in polarity entropy $H(A)$, indicating higher certainty in the remaining predictions;
    \item an increase in sentiment dominance $D(A)$, reflecting more decisive classifications;
    \item shifts in the ranking of top aspects, revealing the stability (or fragility) of salience assessments under stricter confidence regimes.
\end{itemize}

This filtering-based stress test does not involve input perturbation but yields interpretable diagnostics of model behavior in terms of consistency and reliability. For each threshold level, we recompute the soft-count $\tilde{w}(A, s)$, entropy $H(A)$, dominance $D(A)$, and aspect rankings, and compare them across levels to quantify sentiment stability.

\paragraph{Bootstrap Resampling for Stability Assessment.}
While confidence-based filtering assesses robustness to prediction uncertainty, bootstrap resampling evaluates a different dimension: the statistical stability of the indicators themselves. It answers the question: how much would our calculated entropy or dominance scores change if we had collected a slightly different sample of data?

To quantify this, we employ a non-parametric bootstrap procedure \cite{Efron1993}. For each key aspect, we treat its $N_A$ observed occurrences as an empirical distribution. We then generate a large number of bootstrap samples (e.g., $B=1000$) by drawing $N_A$ instances with replacement from this original set. For each bootstrap sample, we recompute the metrics of interest (Polarity Entropy $H(A)$ and Sentiment Dominance $D(A)$).

This process yields an empirical distribution for each indicator, from which we can construct confidence intervals (e.g., 95\% CIs). Specifically, we use the percentile method, where the 95\% confidence interval is determined by the 2.5th and 97.5th percentiles of the bootstrap distribution:

\begin{equation}
CI_{0.95}(\hat{\theta}) = \left[ \hat{\theta}^{*(0.025)}, \; \hat{\theta}^{*(0.975)} \right],
\end{equation}

where $\hat{\theta}^{*(q)}$ denotes the $q$-th quantile of the bootstrap distribution of the statistic $\hat{\theta}$ (e.g., $H(A)$ or $D(A)$).

A narrow confidence interval suggests that the metric is a stable and reliable estimate, largely independent of the specific sample of text collected. Conversely, a wide interval indicates that the metric is highly sensitive to sampling variability and should be interpreted with caution.

\subsection{Alignment with the SAFE Framework}
Our approach extends the applicability of the core principles of SAFE to a previously unexplored setting, namely the evaluation of models in unlabeled-data contexts where standard metrics (e.g., accuracy) cannot be applied.

 The Polarity Entropy and Sentiment Dominance indicators provide a quantitative proxy for decisional clarity, conceptually aligned with the Rank Graduation Explainability (RGE) metric. While RGE quantifies the contribution of input variables by comparing full and reduced models, our method adapts this logic to aspect-level sentiment by analyzing the informativeness of polarity distributions. Similarly, our robustness assessments serve a purpose parallel to Rank Graduation Robustness (RGR). Our contribution, therefore, is to extend some SAFE principles to unstructured text analysis in model-agnostic, data-constrained pipelines.

\section{A case study: Reddit discourse}
\subsection{The data}
The dataset was assembled during February 2025 by employing the Python Reddit API Wrapper (PRAW). It focuses on content sourced from the subreddits r/environment, r/climatechange and r/renewableenergy, which discuss issues such as pollution, climate change and sustainable development \cite{redditAPI2023}.

During the data scraping phase, we retained only posts and comments written in English, while excluding bot-generated content.
In total, we gathered 978 posts and 11,547 comments, then organized them into two separate subsets: one for the primary posts and the other for all corresponding user comments. Each entry in the final corpus contains relevant metadata (post or comment ID, author, timestamp, score, subreddit, flair) as well as the actual text.

We performed a light text preprocessing, removing only noise elements such as missing fields, user mentions, URLs, while preserving punctuation, numerical expressions, stopwords, and word inflections to retain the contextual information required by transformer-based models.

Aspect-based sentiment predictions were generated using PyABSA v2.4.1.post1, an open-source library that offers streamlined functionalities for Aspect Term Extraction and Polarity Classification (ATEPC). We employed the official English ATEPC checkpoint distributed with the library, running inference with default hyperparameters. This process identifies specific topics within the text (e.g., “climate,” “oil”) and assigns a sentiment label (positive, negative, or neutral) to each. Crucially for our analysis, for each aspect occurrence, we retained the complete softmax distribution over {Positive, Negative, Neutral}, which was subsequently aggregated to compute Dominance and Entropy (see Section \ref{par2.2}).

\subsection{Principal Aspects, Polarity and Statistical Indicators}
Overall sentiment distribution differs significantly between original posts and subsequent comments. In the posts, the sentiment classification reveals 11.2\% positive, 48.6\% negative, and 40.1\% neutral. By contrast, comments exhibit a more pronounced negative polarity, with 11.3\% positive, 67.6\% negative, and 21.1\% neutral. One possible explanation for this discrepancy is that original posts often seek to introduce or summarize a topic in a more balanced or neutral manner, whereas user comments may adopt a more critical or skeptical perspective, thereby amplifying the share of negative sentiment.

\begin{figure}[!ht]
    \centering
    \begin{subfigure}{0.45\textwidth}
        \includegraphics[width=\linewidth]{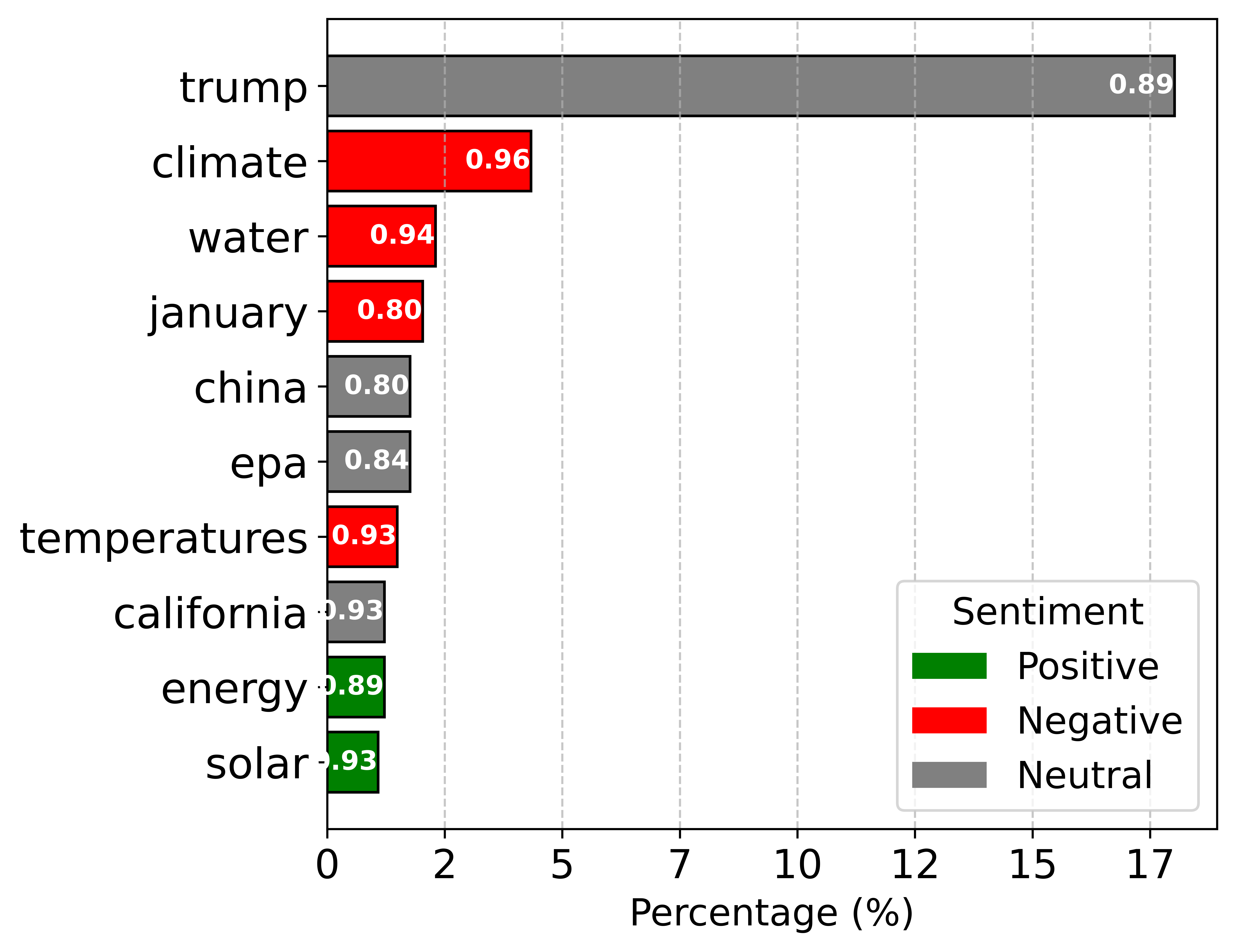} 
        \scriptsize{\caption{Post predictions}}
    \end{subfigure}
    \hfill
    \begin{subfigure}{0.45\textwidth}
        \includegraphics[width=\linewidth]{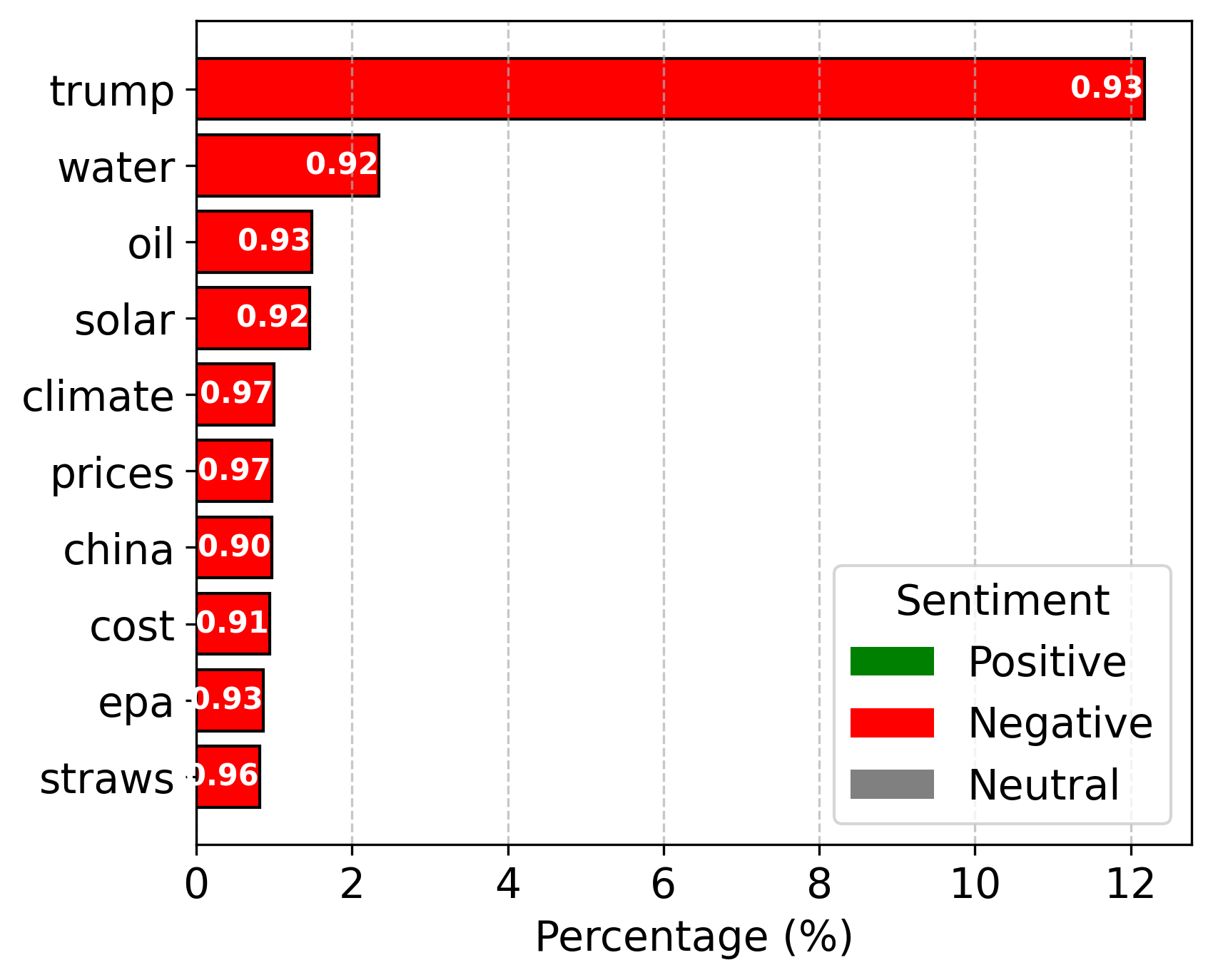} 
        \scriptsize{\caption{Comment predictions}}
    \end{subfigure}

    \vspace{1em} 

    \begin{subfigure}{0.45\textwidth}
        \includegraphics[width=\linewidth]{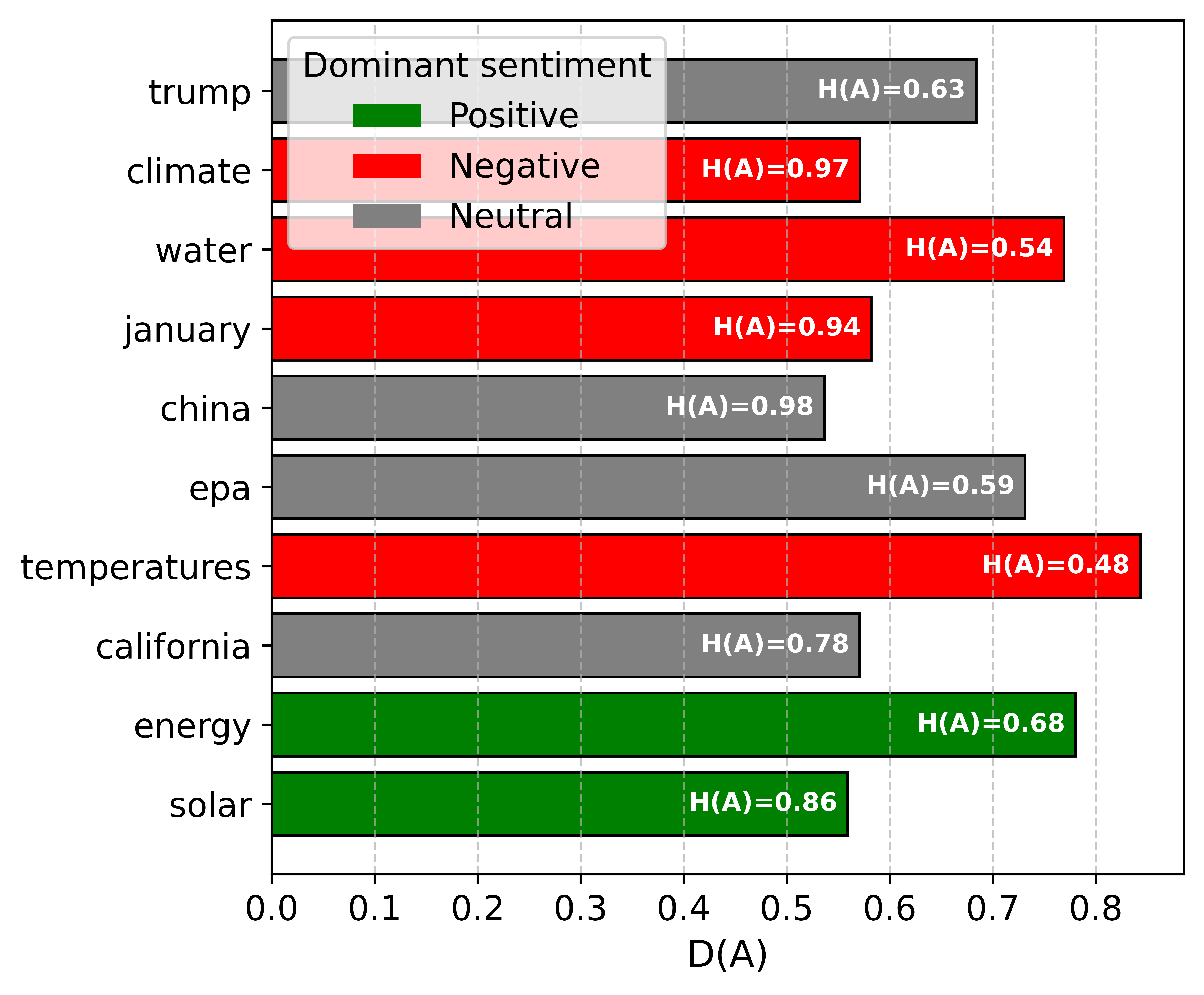} 
        \scriptsize{\caption{Dominance/Entropy in Posts}}
    \end{subfigure}
    \hfill
    \begin{subfigure}{0.45\textwidth}
        \includegraphics[width=\linewidth]{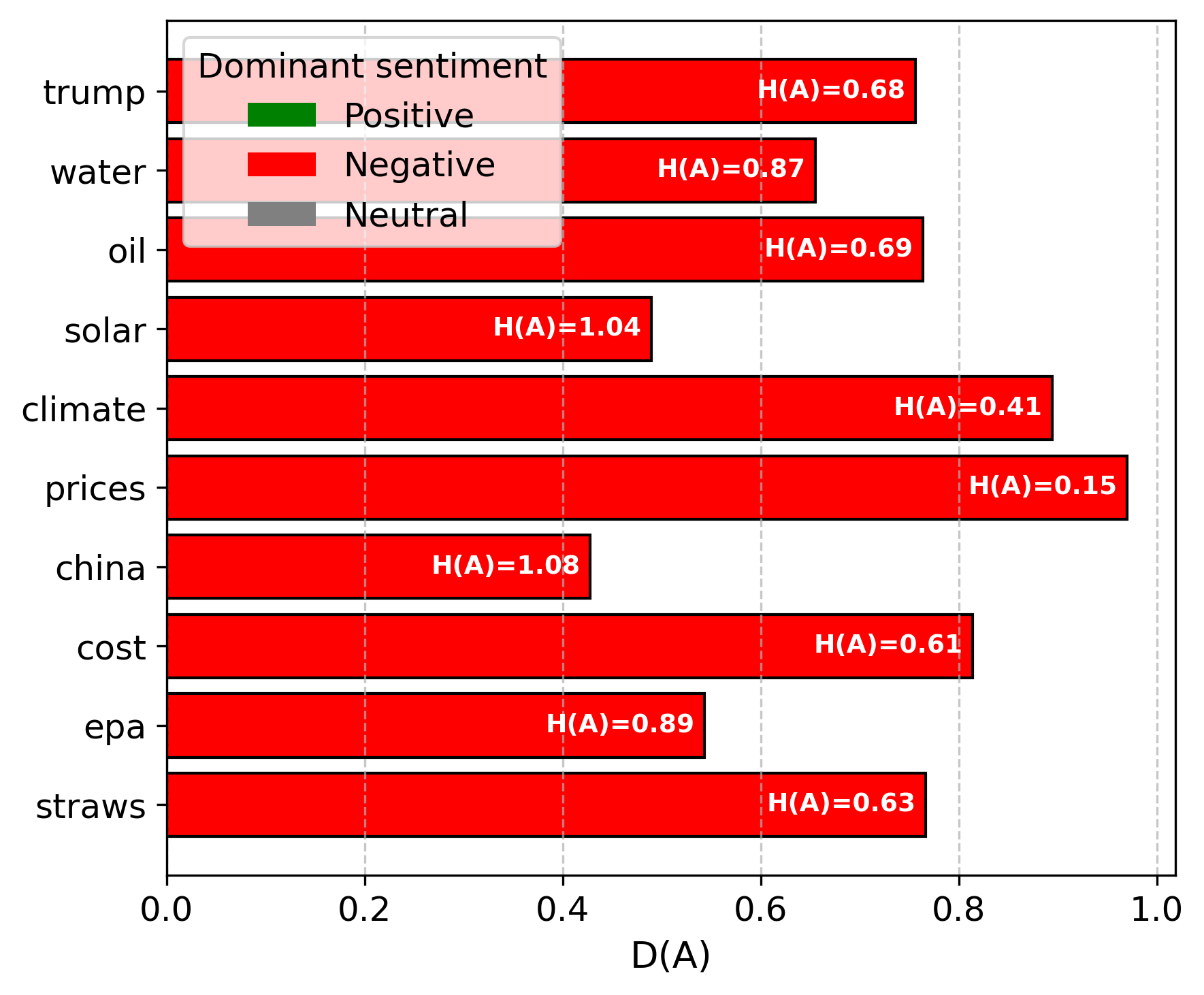} 
        \scriptsize{\caption{Dominance/Entropy in Comments}}
    \end{subfigure}

  \caption{Top 10 ABSA results, Confidence vs Dominance/Entropy}
  \parbox{\linewidth} {\scriptsize{Side-by-side comparison of model outputs for posts and comments. Panels (a-b) report the percentage of the ABSA predictions with Confidence C(A), while panel (c-d) display Dominance $D(A)$ and Entropy $H(A)$ for the same data. }}\label{fig1}
\end{figure}

Figures \ref{fig1}(a-b) provide a classical representation for ABSA, reporting the percentage frequency of the top aspects in posts (a) and comments (b), respectively. The confidence C(A) for each aspect is displayed on its corresponding bar. The color label is given by the most frequent sentiment class. 

Figures \ref{fig1}(c-d) illustrate our proposed metrics (see Section \ref{par2.2}). In these charts, the bar length represents the dominance $D(A)$ for the top aspects in posts (c) and comments (d), respectively. The  entropy $H(A$) is displayed as a label within each bar, and the color is assigned based on the dominant sentiment class. This representation captures the probabilistic structure of the model’s predictions rather than their discrete, frequency-based counts.

The comparison reveals several insights. First, aspects with low entropy (e.g., prices, temperatures) correspond to cases where the model is reasonably certain in its predictions. Conversely, an aspect with high entropy like \emph{china} indicates model uncertainty and a more ambiguous distribution of probabilities. Secondly, we observe that some aspects exhibit a marked discrepancy between
Confidence and Dominance. For example, aspects such as \emph{china} and \emph{solar} show consistently high Confidence values in both datasets, indicating that the model is certain and decisive in its instance-level
predictions, yet low Dominance scores, suggesting that opinions are strongly divided across sentiment classes. This combination---high Confidence coupled
with low Dominance---constitutes the statistical signature of a highly polarizing aspect: while individual predictions are clear and confident, they are distributed across opposing sentiment categories. By contrast, when both Dominance and Confidence are high, as in the case of \emph{prices}, this indicates that the model is not only decisive in its individual predictions
but also that a single sentiment class overwhelmingly prevails at the aggregate level. Such a pattern is indicative of strong consensus on the aspect.

Finally, the comparative table \ref{tab:1} confirms these findings numerically. It highlights that aspects with low entropy are associated with high Dominance, reflecting model certainty, while aspects with higher entropy reveal weaker separation between classes and greater uncertainty.

\renewcommand{\arraystretch}{1.25}
\begin{table}[htbp]
\centering
\scriptsize
\caption{Comparative statistics of top aspects: frequency, sentiment mode, confidence, dominance, dominant label, and  entropy.}
\vspace{0.5em}
\begin{tabular}{l|lcccccc}
\hline
& & & & & &\\
Type & Aspect & \shortstack{Freq \\ (\%)} & \shortstack{Sentiment \\ mode} & \shortstack{$C(A)$} & \shortstack{$D(A)$} & \shortstack{$D(A)$ \\ label} & \shortstack{$H(A)$} \\

\hline

& trump         & 18.02 & Neutral  & 0.89 & 0.68 & Neutral  & 0.63 \\
& climate       &  4.34 & Negative & 0.96 & 0.57 & Negative & 0.97 \\
& water         &  2.30 & Negative & 0.94 & 0.77 & Negative & 0.54 \\
& january       &  2.03 & Negative & 0.80 & 0.58 & Negative & 0.94 \\
Posts & china         &  1.76 & Neutral  & 0.80 & 0.54 & Neutral  & 0.98 \\
& epa           &  1.76 & Neutral  & 0.84 & 0.73 & Neutral  & 0.59 \\
& temperatures  &  1.49 & Negative & 0.93 & 0.84 & Negative & 0.48 \\
& california    &  1.22 & Neutral  & 0.93 & 0.57 & Neutral  & 0.78 \\
& energy        &  1.22 & Positive & 0.89 & 0.78 & Positive & 0.68 \\
& solar         &  1.08 & Positive & 0.93 & 0.56 & Positive & 0.86 \\
\hline

& trump    & 12.18 & Negative & 0.93 & 0.76 & Negative & 0.68 \\
& water    &  2.35 & Negative & 0.92 & 0.66 & Negative & 0.87 \\
& oil      &  1.48 & Negative & 0.93 & 0.76 & Negative & 0.69 \\
& solar    &  1.46 & Negative & 0.92 & 0.49 & Negative & 1.04 \\
Comments & climate  &  1.00 & Negative & 0.97 & 0.89 & Negative & 0.41 \\
& prices   &  0.97 & Negative & 0.97 & 0.97 & Negative & 0.15 \\
& china    &  0.97 & Negative & 0.90 & 0.43 & Negative & 1.08 \\
& cost     &  0.94 & Negative & 0.91 & 0.81 & Negative & 0.61 \\
& epa      &  0.86 & Negative & 0.93 & 0.54 & Negative & 0.89 \\
& straws   &  0.81 & Negative & 0.96 & 0.77 & Negative & 0.63 \\
\hline
\end{tabular}
\label{tab:1}
\end{table}

\paragraph{JSD Divergence: Posts vs. Comments.}

To measure internal debate within a community, we compare the sentiment profile of original posts ($\tilde{w}_{\textit{posts}}$) with that of the corresponding comments ($\tilde{w}_{\textit{comments}}$). A low Jensen–Shannon divergence (JSD) score indicates consensual discourse, whereas a high score reveals controversy, suggesting that the community's reaction differs significantly from the initial statement.

\renewcommand{\arraystretch}{1.25}
\begin{table}[ht!]
\centering
\scriptsize
\caption{JSD between posts and comments for the most frequent shared aspects}\label{tab:2}
\vspace{0.5em}
\begin{tabular}{lc|cc|cc|cc|p{4.6cm}}
\hline
Aspect & JSD
       & \multicolumn{2}{c|}{$D(A)$-labels}
       & \multicolumn{2}{c|}{$D(A)$}
       & \multicolumn{2}{c|}{$H(A)^*$}
       & Note \\
\cline{3-4}\cline{5-6}\cline{7-8}
      &                     & Posts & Comm. & Posts & Comm. & Posts & Comm. & \\
\hline
trump       & 0.06 & \textbf{Neu} & \textbf{Neg} & 0.73 & 0.52 & 0.59 & 0.72 & Marginal polarity shift; increased uncertainty in comm. (\(D\downarrow\), \(H\uparrow\))\\
climate     & 0.07 & Neg & Neg & 0.63 & 0.89 & 0.83 & 0.37 & Low divergence; \emph{greater certainty in comments} (\(D\uparrow, H\downarrow\))\\
temperature & 0.06 & Neg & Neg & 0.75 & 0.95 & 0.67 & 0.22 & Low divergence; \emph{greater certainty in comments} (\(D\uparrow, H\downarrow\))\\
solar       & \textbf{0.25} & Neg & Neg & 0.66 & 0.43 & 0.71 & 0.93 & Medium divergence driven by \emph{flattening toward neutrality} in comments (\(D\downarrow\), \(H\uparrow\))\\
china       & 0.04 & Neu & Neu & 0.38 & 0.46 & 0.99 & 0.84 & Very low divergence; profiles nearly unchanged with \emph{persistently high uncertainty} on both sides (high \(H\))\\
water       & 0.03 & Neg & Neg & 0.87 & 0.83 & 0.35 & 0.53 & Very low divergence; posts and comments exhibit quite similar profiles, both with high \(D\) and low \(H\) (slightly less certainty in comments)\\
price       & 0.09 & Neg & Neg & 0.80 & 0.94 & 0.53 & 0.21 & Low divergence; slightly more certainty in comments (lower \(H\))\\
oil         & 0.07 & Neg & Neg & 0.99 & 0.83 & 0.52 & 0.51 & Low divergence; quite similar profiles\\
battery     & \textbf{0.32} & \textbf{Pos} & \textbf{Neg} & 0.75 & 0.73 & 0.52 & 0.70 & High divergence due to \emph{polarity flip} accompanied by \emph{greater uncertainty in comments} (\(H\uparrow\))\\
fuel        & \textbf{0.25} & Neg & Neg & 0.54 & 0.88 & 0.71 & 0.36 & Medium divergence from \emph{much greater certainty in comments} (strong \(D\uparrow, H\downarrow\))\\
\hline
\end{tabular}
\vspace{0.9em}
\parbox{\linewidth}{\scriptsize \textit{*} Entropy is reported without normalization. Since the number of sentiment classes is fixed to three,
$H(A) \in [0, \ln 3]$. Normalized entropy can be obtained as $H_{norm}(A) = {H(A)} / {\ln 3}$
which is a monotone rescaling that leaves all comparisons and interpretations unchanged.}
\end{table}

\vspace{0.8em}

Table  \ref{tab:2} illustrates how the JSD quantifies the distance between the sentiment profiles produced by ABSA on posts and comments, while Dominance D(A) and Entropy H(A) make explicit the mechanisms underlying such distance. In particular:
\begin{itemize}
  \item \emph{Medium-high divergences} emerge when: (i) a polarity flip occurs (i.e., a change in the dominant sentiment class); or (ii) — conditional on the same dominant class — there is a substantial redistribution of probability mass across classes (e.g., strong drift toward neutrality or strong concentration on the dominant class). 
  \item \emph{Low divergences}, by contrast, indicate nearly unchanged profiles, including cases where both channels exhibit high uncertainty (high \(H\)).
\end{itemize}

\subsection{Robustness and Stability Assessment}

\begin{figure}[!ht]
    \centering
   \includegraphics[width=0.85\linewidth]{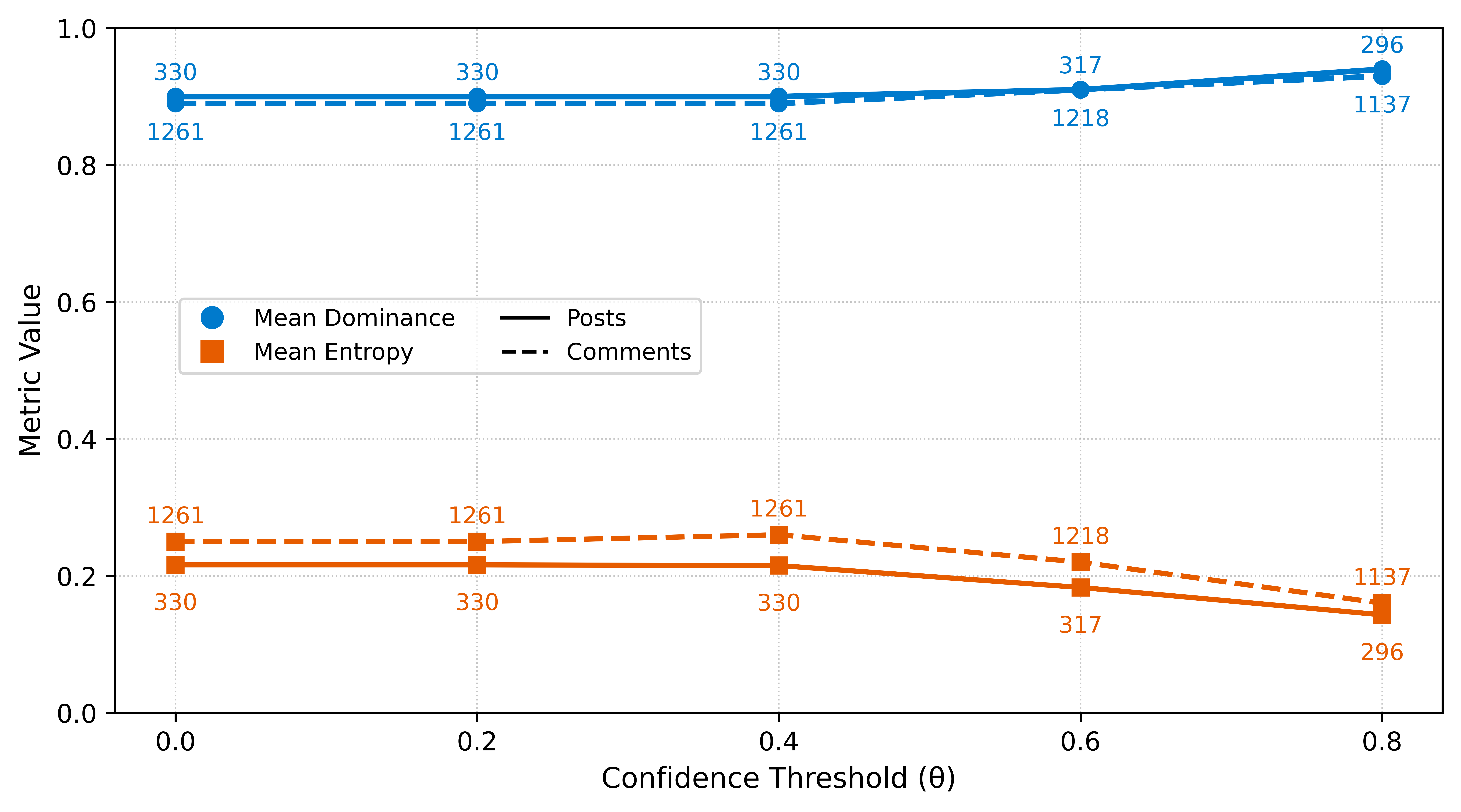}
  \caption{Explainability-Robustness Tradeoff via Confidence Filtering}\label{fig2}
\end{figure}

To assess the robustness and stability of our proposed indicators, we evaluate how entropy and dominance respond to confidence-based filtering. Figure~\ref{fig2} illustrates this relationship for both posts and comments.

As the confidence threshold $\theta$ increases, the \emph{Mean Entropy}, calculated on the filtered set of aspects, decreases monotonically. This indicates that the retained predictions become less ambiguous and more homogeneous. Conversely, the \emph{Mean Dominance} increases, reflecting sharper classifications and a stronger prevalence of the modal sentiment class.

The nearly parallel trajectories for posts and comments confirm that these indicators behave consistently across both data sources. However, comments consistently exhibit slightly higher entropy than posts at comparable thresholds, suggesting that user interactions are inherently more diverse in sentiment. Nonetheless, both converge toward higher certainty and clearer dominance under stricter filtering conditions.

Figure~\ref{fig2} also reports the number of unique aspects preserved after filtering, with counts annotated alongside their respective lines (solid for posts, dashed for comments). These counts decline only marginally with stricter thresholds—from 330 to 296 in posts, and from 1261 to 1137 in comments—indicating that the model’s expressive capacity remains largely intact, even though the retained instances are fewer yet more reliable.

We next assess the robustness of the proposed explainability indicators—$\tilde{w}(A)$, dominance $D(A)$, and entropy $H(A)$—through non-parametric bootstrap resampling (par.~\ref{par:2.3}). For the top-10 aspects retained at $\theta=0.8$, we generated 1000 bootstrap replicates and recomputed the indicators to evaluate their stability under sampling variability.

Table~\ref{tab:3} reports the mean values together with 95\% confidence intervals. The results show that stability is closely aligned with both frequency and the entropy–dominance profile of each aspect. Frequent aspects such as \emph{trump} yield narrow confidence intervals, indicating that both $H(A)$ and $D(A)$ are robust to resampling. By contrast, less frequent aspects (e.g., \emph{straw}, \emph{oil}, \emph{plastic}) exhibit wider intervals, reflecting greater sensitivity to sample fluctuations.

Equally important, the bootstrap confirms the interpretive distinction captured by our indicators: aspects with low entropy and high dominance (e.g., \emph{price}, \emph{temperature}) consistently display strong consensus and stable sentiment profiles, while aspects with high entropy and low dominance (e.g., \emph{china}, \emph{trump}) are less stable and indicative of polarized discourse.

Overall, these findings validate that the proposed indicators are not only theoretically sound but also empirically robust, providing a reliable basis for explainable analysis of ABSA outputs.

\renewcommand{\arraystretch}{1.2}
\begin{table}[!ht]
\centering
\caption{Bootstrap stability of top-10 aspects at confidence threshold $\theta=0.8$.}
\parbox{\linewidth}{\scriptsize For each aspect, $N$ is the number of occurrences, $H(A)$ the entropy with 95\% CI, and $D(A)$ the dominance with 95\% CI.}
\vspace{0.5em}
\label{tab:3}
\begin{tabular}{l|lcccc}
\hline
Type & Aspect & $N$ & $H(A)$ [95\% CI] & $D(A)$ [95\% CI] \\
\hline
 & trump       & 118 & 0.54 [0.43, 0.65] & 0.78 [0.70, 0.85] \\
 & climate     &  22 & 0.78 [0.46, 0.97] & 0.63 [0.45, 0.85] \\
 & temperature &  18 & 0.61 [0.27, 0.87] & 0.76 [0.59, 0.93] \\
 & solar       &  12 & 0.56 [0.07, 0.89] & 0.76 [0.53, 0.99] \\
Posts & china       &  10 & 0.91 [0.63, 1.00] & 0.48 [0.36, 0.68] \\
 & water       &   9 & 0.07 [0.01, 0.14] & 0.98 [0.96, 1.00] \\
 & forest      &   7 & 0.56 [0.36, 0.63] & 0.66 [0.54, 0.86] \\
 & epa         &   7 & 0.56 [0.36, 0.63] & 0.65 [0.56, 0.86] \\
 & plastic     &   6 & 0.33 [0.02, 0.65] & 0.84 [0.50, 1.00] \\
 & germany     &   6 & 0.09 [0.01, 0.22] & 0.98 [0.94, 1.00] \\
\hline
 & trump        & 388 & 0.71 [0.67, 0.75] & 0.53 [0.49, 0.57] \\
 & price        &  66 & 0.19 [0.05, 0.32] & 0.95 [0.89, 0.99] \\
 & water        &  56 & 0.39 [0.19, 0.58] & 0.88 [0.80, 0.95] \\
 & china        &  49 & 0.82 [0.69, 0.93] & 0.51 [0.43, 0.61] \\
Comments & climate      &  45 & 0.32 [0.11, 0.53] & 0.90 [0.82, 0.97] \\
 & oil          &  38 & 0.40 [0.18, 0.62] & 0.88 [0.78, 0.96] \\
 & temperature  &  36 & 0.14 [0.03, 0.28] & 0.97 [0.92, 0.99] \\
 & fuel         &  34 & 0.30 [0.07, 0.48] & 0.91 [0.81, 0.99] \\
 & straw        &  30 & 0.37 [0.13, 0.56] & 0.88 [0.75, 0.97] \\
 & battery      &  29 & 0.37 [0.16, 0.54] & 0.86 [0.73, 0.96] \\
\hline
\end{tabular}
\end{table}

\vspace{0.8em}

\section{Discussion and Conclusions}

This work proposed a statistical, model-agnostic framework for explaining the behaviour of pretrained ABSA models on unlabeled text. Rather than relying on
token-level post-hoc explainers or accuracy on annotated benchmarks, we characterised model behaviour through indicators computed on
soft-count probability outputs: aspect-level entropy $H(A)$, dominance $D(A)$, and Jensen–Shannon divergence $JSD(A)$ across sources. The case study on
environmental discourse in Reddit communities demonstrates that these quantities provide complementary, interpretable diagnostics of certainty,
decisiveness, and cross-source shifts in sentiment.

\paragraph{Key empirical findings.}
The empirical analysis of Reddit discourse highlights several consistent patterns. Confidence-based filtering induces a monotonic decrease in aspect entropy and a corresponding increase in aspect dominance, reflecting the expected trade-off between coverage and decisional clarity. The Jensen-Shannon Divergence isolates where sentiment profiles differ across conversational roles (posts vs. comments), while entropy and dominance quantify whether this divergence is driven by shifts in polarity or by changes in decisiveness. Furthermore, bootstrap resampling shows that the uncertainty of these indicators is itself measurable: frequent aspects yield narrow confidence intervals, while rarer ones display wider ranges, making the link between sample size and inferential stability explicit.

\paragraph{Statistical contribution.}
From a statistical perspective, our framework's primary contribution is demonstrating how simple, model-agnostic functionals of softmax outputs can be reframed as robust estimators with well-defined properties. Specifically, we interpret aspect entropy $H(A)$ as a measure of dispersion, dominance $D(A)$ as a maximum-probability functional, and the Jensen-Shannon Divergence JSD(A) as a bounded divergence between empirical distributions.

Crucially, their behavior under resampling allows for a direct assessment of their stability, providing insights into both bias (systematic uncertainty) and variance (sensitivity to sampling). This yields a transparent diagnostic toolkit that can be applied to characterize the behavior of any underlying model architecture.

\paragraph{Implications.}
The proposed measures support three practical aims. First, they enable pre-deployment validation: wide bootstrap intervals or high divergence values
signal potential instability before annotation resources are committed.
Second, they guide annotation by identifying aspects with high entropy or high divergence, i.e.\ those most informative for model improvement. Third, they
offer a mechanism for post-deployment monitoring: systematic shifts in these statistics over time may serve as proxy indicators of model drift. In each
case, the indicators act as statistical summaries that are computationally cheap, interpretable, and comparable across datasets.

\paragraph{Limitations and future works.}
The proposed indicators are model-agnostic and retain a consistent interpretative meaning: entropy always quantifies uncertainty, dominance
captures decisiveness, and divergence highlights cross-source shifts.
However, their empirical values necessarily depend on the corpus on which they are applied and on the reliability of aspect extraction.
Aspects with very few occurrences may yield wide bootstrap confidence intervals, reflecting sensitivity to sample size rather than instability
of the indicator itself. Similarly, different extraction pipelines may identify partially different aspect sets, leading to shifts in the profiles
to which the indicators are applied. These considerations do not undermine the validity of the measures, but clarify that their informativeness is
conditional on the representativeness of the underlying data.

For future works, alternative aggregation schemes for $\tilde{w}(A,s)$, and extensions to other measures of dispersion (e.g., Rényi
entropy) or distance (e.g., Hellinger divergence), warrant exploration.

\paragraph{Conclusion.}
Statistical indicators computed on soft outputs offer a simple, transparent and computationally efficient route to explain ABSA model behaviour in the wild.
By combining entropy, dominance and divergence with confidence filtering and bootstrap quantification, we obtain interpretable evidence on certainty,
decisiveness and stability—without labels or model internals. This provides a practical path toward safer, more auditable deployments of ABSA systems and a
scalable complement to conventional XAI tools.

\end{document}